\documentclass[
preprint,
nofootinbib,
amsmath,amssymb,
aps,
]{revtex4-2}

\usepackage{graphicx}
\usepackage{dcolumn}
\usepackage{multirow}
\usepackage{bm}
\usepackage{siunitx}
\usepackage{hyperref}
\usepackage{ulem}
\usepackage{comment}

\begin{document}

\title{Determination of compressive stress in thin films using micro-machined buckled membranes}

\author{C. Malhaire}
\email{christophe.malhaire@insa-lyon.fr}
\affiliation{Univ Lyon, INSA Lyon, CNRS, Ecole Centrale de Lyon, Universit\'e Claude Bernard Lyon 1, CPE Lyon, INL, UMR5270, 69621 Villeurbanne, France}

\author{M. Granata}
\author{D. Hofman}
\affiliation{Laboratoire des Matériaux Avanc\'es-IP2I, CNRS, Universit\'e de Lyon, Universit\'e Claude Bernard Lyon 1, F-69622 Villeurbanne, France}

\author{A. Amato}
\altaffiliation[Current affiliation: ]{Maastricht University, P.O. Box 616, 6200 MD Maastricht, The Netherlands}
\author{V. Martinez}
\author{G. Cagnoli}
\affiliation{Universit\'e de Lyon, Universit\'e Claude Bernard Lyon 1, CNRS, Institut Lumi\`ere Mati\`ere, F-69622 Villeurbanne, France}

\author{A. Lemaitre}
\affiliation{Navier, Ecole des Ponts, Universit\'e Gustave Eiffel, CNRS, Marne-la-Vall\'ee, France}

\author{N. Shcheblanov}
\affiliation{MSME, UMR 8208, Universit\'e Gustave Eiffel, CNRS, Universit\'e Paris-Est Cr\'eteil, Marne-la-Vall\'ee, France}

\date{\today}

\begin{abstract}
In this work, optical profilometry and finite-element simulations are applied on buckled micro-machined membranes for the stress analysis of ion-beam-sputtered $\mathrm{Ta_{2}O_{5}}$ and $\mathrm{SiO_{2}}$ thin films. Layers with different thicknesses are grown on silicon substrates, then several membranes with different geometries are manufactured with standard micro-system technologies; due to a high level of the films' compressive stress, buckled membranes are obtained. Thermally-grown silica membranes are also produced, for comparison. The residual stress values are determined by comparing the measured and simulated deflections of the membranes. The average stress state of the $\mathrm{Ta_{2}O_{5}}$ thin films is found to be $-209$ MPa. The $\mathrm{SiO_{2}}$ thin films are in a higher compressive stress state whose average value is $-576$ MPa. The average stress in thermal $\mathrm{SiO_{2}}$ thin layers grown at 1130 $^{\circ}$C is found equal to $-321$ MPa, in good agreement with the literature.
\end{abstract}

\maketitle

\section{Introduction}
The current high-reflection (HR) coatings of the Advanced LIGO, Advanced Virgo and KAGRA gravitational-wave detectors \cite{aLIGO, AdVirgo, KAGRA} are thickness-optimized Bragg reflectors \cite{Villar10} of alternating layers of ion-beam-sputtered (IBS) tantalum pentoxide (Ta$_2$O$_5$, also known as {\it tantala}, with high refractive index $n_{\textrm{H}}$) and silicon dioxide (SiO$_2$, {\it silica}, low refractive index $n_{\textrm{L}}$), grown by the Laboratoire des Mat\'{e}riaux Avanc\'{e}s\footnote{lma.in2p3.fr} (LMA) \cite{Pinard17, Degallaix19}. Despite their excellent optical and mechanical properties \cite{Degallaix19, Amato19, Granata20}, their Brownian thermal noise \cite{saulson_book} constitutes a severe limitation to the sensitivity of current and future gravitational-wave detectors. Thus, in the last two decades, a considerable research effort has been committed to identify and optimize alternative coating materials of even lower mechanical and optical losses \cite{Granata20_CTN_review,Vajente21}. The need to develop alternative coating materials is even more critical for cryogenic gravitational-wave detectors, either present or future, such as KAGRA, Einstein Telescope \cite{ET1, ET2}, and Cosmic Explorer \cite{Abbott17}. The accurate estimation of residual stress in thin films falls within this context.

In the past, many studies were conducted to determine the stress of tantala and silica coatings. For both materials, the stress is generally compressive (here, negative by convention).

Concerning tantala, a stress value of $-250~\si{\mega\pascal}$ was reported by Ngaruiya \textit{et al.} for $100~\si{\nano\meter}$ thick films grown via reactive DC magnetron sputtering \cite{Ngaruiya_2003}; more precisely, they observed that the stress decreases with increasing oxygen flow rate and becomes almost constant at $-250~\si{\mega\pascal}$ for an oxygen flow greater than 6 sccm. For a $4800~\si{\nano\meter}$ thick film grown via RF magnetron sputtering, Cheng \textit{et al.} measured stress values ranging from $-314~\si{\mega\pascal}$ at 25 $^{\circ}$C to $-346~\si{\mega\pascal}$ at 100 $^{\circ}$C \cite{Cheng_1999}. In a work by Farhan \textit{et al.} on $200~\si{\nano\meter}$ to $350~\si{\nano\meter}$ thick films grown by ion beam-assisted deposition, the stress was found to increase with the ion current density from $-280~\si{\mega\pascal}$ to $-375~\si{\mega\pascal}$ \cite{Farhan_2013}.

Concerning silica, films of high optical quality with stress ranging from $-490~\si{\mega\pascal}$ to $-48~\si{\mega\pascal}$ were obtained by IBS, using high-energy $\mathrm{O}_{2}$ ion bombardment during deposition \cite{Davenport_2020}.

In 2005, the stress level of IBS silica and tantala thin layers in the HR coatings of the LIGO interferometer was determined by Netterfield \textit{et al.} before and after a post-deposition thermal treatment ({\it annealing} hereafter) was applied \cite{Netterfield_2005}. They showed that such tantala films can readily go from a compressive to a tensile state after only moderate annealing, whereas the stress state in silica is only slightly relaxed and remains compressive. It was also pointed out that for multi-layer HR coatings composed of several silica/tantala doublets, the net post-anneal stress depends on the silica to tantala thickness ratio.

Netterfield \textit{et al.} also argued that different annealing regimes adopted by various coating suppliers might lead to quite large variations in the residual stress, thus explaining the differences in mechanical loss observed for otherwise similar multi-layer coatings. More recently, Granata \textit{et al.} showed that the measured coating dissipation in HR multi-layer coatings is usually higher than the expectation \cite{Granata20,Granata16}, and that their stress level could be loosely correlated with the measured excess loss \cite{Granata16}.

Finally, it is known that residual stresses can promote the appearance of cracks during annealing  \cite{Abernathy_2021,Lalande_2021,Lalande_master_thesis_2021}.

In all the works cited above, the substrate curvature method associated with the Stoney's formula was employed to quantify the stress state of the films. Indeed, this is a well-known, non destructive method used to determine the stress level of a thin film on a thick substrate, sometimes even in the case where the hypotheses of the model are not completely satisfied \cite{Ardigo_2014}. However, apart from the assumptions of the model, this method suffers from a few practical flaws: dependence on the substrate parameters (thickness non-uniformity is critical, as it is squared in the Stoney's equation), and repeatability of measurements (a three-point support is essential). Therefore, it is generally only reliable inasmuch as it is used in differential measurements (before and after annealing, for instance) rather than to estimate absolute stress values.

In this paper, we present an alternative method for the determination of compressive stresses in thin films. It consists of manufacturing buckled membranes by locally etching the substrate under the thin films and of comparing the measured deflection profiles of the membranes with those obtained by finite element simulations. In the case of thin films with residual tensile stress, from which flat membranes would be obtained, other characterization techniques such as vibrometry \cite{Malhaire_RSI_2012}, bulge-test \cite{Martins_Micro_Tech_2009} or point-deflection \cite{Martins_EPJAP_2009} methods should be used.

\section{Experiment}

\subsection{Samples}
After a standard cleaning procedure, 50-mm diameter, double-side polished, silicon wafers were thermally oxidized on both sides (1.5 µm thick films grown at 1130°C in wet atmosphere). The backside $\mathrm{SiO_{2}}$ layer was patterned to open square windows of various dimensions. On the front side, the oxide layer was either left pristine, in order to study the stress state of the thermal $\mathrm{SiO_{2}}$ for comparison, or completely removed before the deposition of thin films.
\begin{table*}
\begin{center}
\caption{\label{tab:membranes_characteristics}Measured (thickness, size, deflection) and estimated (stress) properties of self-supporting thin membranes.}
\begin{ruledtabular}
\begin{tabular}{c|ccccc}
Material & Sample ID & Thickness & Side length & Measured deflection & Stress\\
 &  & (nm) &  (µm) &  (µm) & (MPa)\\
\colrule
\multirow{7}*{IBS $\mathrm{Ta_{2}O_{5}}$} & S20061-07 & 1008 & $390.1 \pm 2.0$ & $-9.5 \pm 0.1$ & $-206$\\
 & S20061-09 & 1008 & $395.4 \pm 2.0$ & $-9.5 \pm 0.1$ & $-206$\\
 & S20061-14 & 1008 & $297.0 \pm 2.0$ & $-7.3 \pm 0.1$ & $-208$\\
 & S20061-06 & 1008 & $193.5 \pm 2.0$ & $-4.7 \pm 0.1$ & $-210$\\
 & S20061-15 & 1008 & $194.6 \pm 2.0$ & $-4.7 \pm 0.1$ & $-210$\\
 & S20196-01 & 1806 & $373.3 \pm 2.0$ & $-9.3 \pm 0.1$ & $-210$\\
 & S20196-03 & 1806 & $276.8 \pm 2.0$ & $-6.9 \pm 0.1$ & $-210$\\
 & S20196-10 & 1806 & $178.3 \pm 2.0$ & $-4.0 \pm 0.1$ & $-211$\\
\colrule
\multirow{4}*{IBS $\mathrm{SiO_{2}}$} & S21056-02 & 1000 & $238.5 \pm 2.0$ & $-13.1 \pm 0.1$ & $-571$ \\
 & S21056-07 & 1000 & $137.5 \pm 2.0$ & $-7.9 \pm 0.1$ & $-577$ \\
 & S22073-01 & 2032 & $246.3 \pm 2.0$ & $-13.7 \pm 0.1$ & $-576$ \\
 & S22073-03 & 2032 & $136.4 \pm 2.0$ & $-7.7 \pm 0.1$ & $-580$ \\
\colrule
\multirow{3}*{Thermal $\mathrm{SiO_{2}}$} & INL-A31 & 1500 & $256.7 \pm 2.0$ & $-11.3 \pm 0.1$ & $-319$ \\
 & INL-C05 & 1500 & $340.5 \pm 2.0$ & $-14.2 \pm 0.1$ & $-320$ \\
 & INL-A21 & 1500 & $506.7 \pm 2.0$ & $-20.9 \pm 0.1$ & $-323$ \\
\end{tabular}
\end{ruledtabular}
\end{center}
\end{table*}

IBS Ta$_2$O$_5$ and SiO$_2$ thin films were grown by the LMA, in a commercial Veeco SPECTOR system, using accelerated, neutralized argon ions as sputtering particles. Prior to deposition, the base pressure inside the vacuum chamber was lower than $10^{-5}$ mbar. Argon was fed into the ion-beam source while oxygen was fed into the vacuum chamber, for a total residual working pressure of the order of $10^{-4}$ mbar. The energy and current of the sputtering ions were 1.3 keV and 0.6 A, respectively.

\begin{figure*}
\begin{center}
\includegraphics[scale = 0.9]{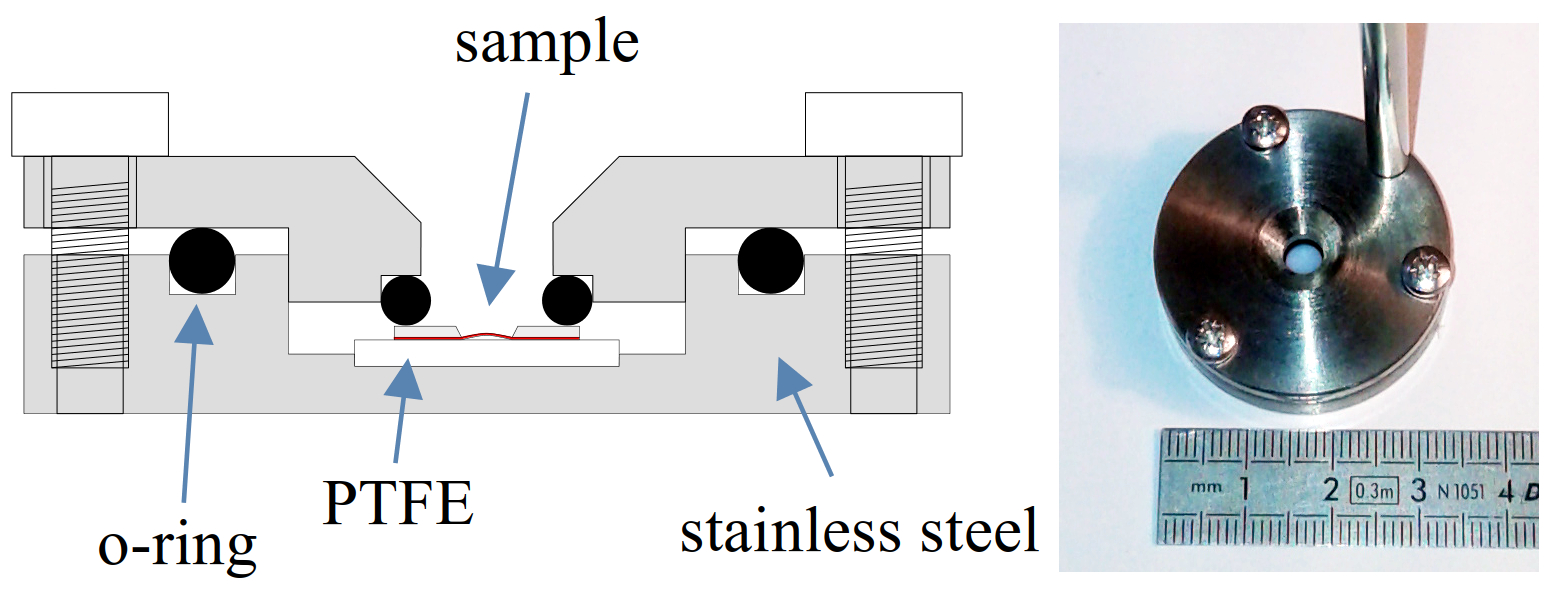}
\caption{\label{fig:fig_sample_holder.jpg}Schematic view and photo of a single-chip sample holder.}
\end{center}
\end{figure*}

After coating deposition, the silicon substrates were etched from their back side windows, through the oxide mask, using an aqueous potassium hydroxide solution (KOH, 34 wt\% at 60 $^{\circ}$C) until self-supporting membranes were obtained out of the thin films. The etching process was generally performed in two stages: a first long etching step, leaving only a few tens of microns of silicon, after which the substrate was cut, so as to complete the etching of each membrane chip individually. During the hour-long etch process, the front side of the whole wafer was protected from KOH by means of a sample holder. A schematic view and a photo of a single-chip sample holder are shown in Fig. \ref{fig:fig_sample_holder.jpg}; a larger sample holder was used for the whole wafer.

After a few tests, the membrane dimensions (side, thickness) were determined so as to guarantee that the compressive stress in the material caused moderate buckling~\cite{Ziebart_1999}, thereby avoiding their rupture during the fabrication process, and obtain reproducible deformed profiles. Indeed, for very thin membranes, the slightest imperfection in their geometry (due to a misalignment of the photolithography mask, over-etching, thickness inhomogeneities, etc.) might lead to a profile that is not symmetrical enough to be correctly measured and fit with finite-element simulations. That was the case for two 250 nm thick SiO$_2$ membranes (140 µm side), for instance, which were so strongly and randomly deformed that they could not be used. So, as a general rule, the more stressed and thinner the film, the less the chance of obtaining membranes with an usable profile. Here, we present results from membranes with thicknesses ranging from about 1 to 2 µm.

After dicing, the 3D profile of each membrane was measured by means of a Wyko NT-1100 white-light profilometer (that is, an interferometric microscope). The membranes were systematically oriented downwards due to the presence of the sample holder during etching: since the sample was sandwiched and clamped against a support, the deformation of the membrane would only occur in one direction.

Samples characteristics are summarized in Table \ref{tab:membranes_characteristics}. The measured deflection refers to the maximum deflection measured at the center of the samples. Note that Table \ref{tab:membranes_characteristics} only lists a few typical samples with mostly distinct geometries, whereas several other similar samples were produced. The results are very reproducible: for example, five $\mathrm{SiO_{2}}$-S21056 membranes (not shown) having the same side size within a 0.75\% uncertainty featured the same stress-induced deflection within a 1.0\% uncertainty. Likewise, the profiles of samples $\mathrm{Ta_{2}O_{5}}$-S20061-6 and -15, as well as samples $\mathrm{Ta_{2}O_{5}}$-S20061-7 and -9 are very similar.

\subsection{Methods}
The refractive index and thickness of the IBS thin films were measured via transmission spectrophotometry through samples grown on fused-silica witness samples. Using {\color{red}a} Perkin Elmer Lambda 1050 spectrophotometer, the spectra were acquired at normal incidence in the 400–1400 nm wavelength range. The refractive index and thickness were first evaluated using the envelope method \cite{Cisneros98}, then those results were used as initial values in a numerical least-square regression analysis. In the model, the adjustable parameters were the film thickness and the ($B_i$, $C_i$) coefficients of the Sellmeier dispersion equation,
\begin{equation}\label{eqn_sellmeier}
n^2 = 1 + \sum^3_{i=1} \frac{B_i \lambda^2}{\lambda^2 - C_i} ,
\end{equation}
where $n$ is the thin film refractive index and $\lambda$ is the wavelength.

Finite-element simulations were performed using ANSYS 2022R1 software. The meshing of the micro-machined membrane was carried out either with 2D (SHELL281) or 3D (SOLID186) elements. In both cases, for a given stress value, the simulated profiles were similar. As the thickness/width ratio of the membranes was small, shell elements were the most effective choice in terms of computation time; the silicon surrounding the membrane was meshed with the 3D elements. In a first static analysis, a very low pressure was applied to the unstressed membrane, in order to obtain a very small downward pre-deflection; then, the geometry of the model was updated. By initially introducing a slight imperfection in the model, it was possible to subsequently obtain a downward buckling of the membrane, as observed experimentally. Finally, a nonlinear buckling analysis was performed by defining a compressive bi-axial stress in the material.

In all simulations, the values of Young's modulus and Poisson's ratio of the membranes were respectively set to $121$ GPa and $0.29$ for tantala samples, and to $78$ GPa and $0.14$ for silica samples, as previously measured on IBS thin films grown under identical conditions \cite{Granata20}; for thermally-grown silica, the values of bulk fused silica were used: $72.8$ GPa and $0.165$, respectively. The dependency of the estimated stress values on these two parameters will be discussed in Section \ref{SECT_res_discuss}.

\section{Results and discussion}
\label{SECT_res_discuss}

\begin{figure}
\begin{center}
\includegraphics[scale = 0.8]{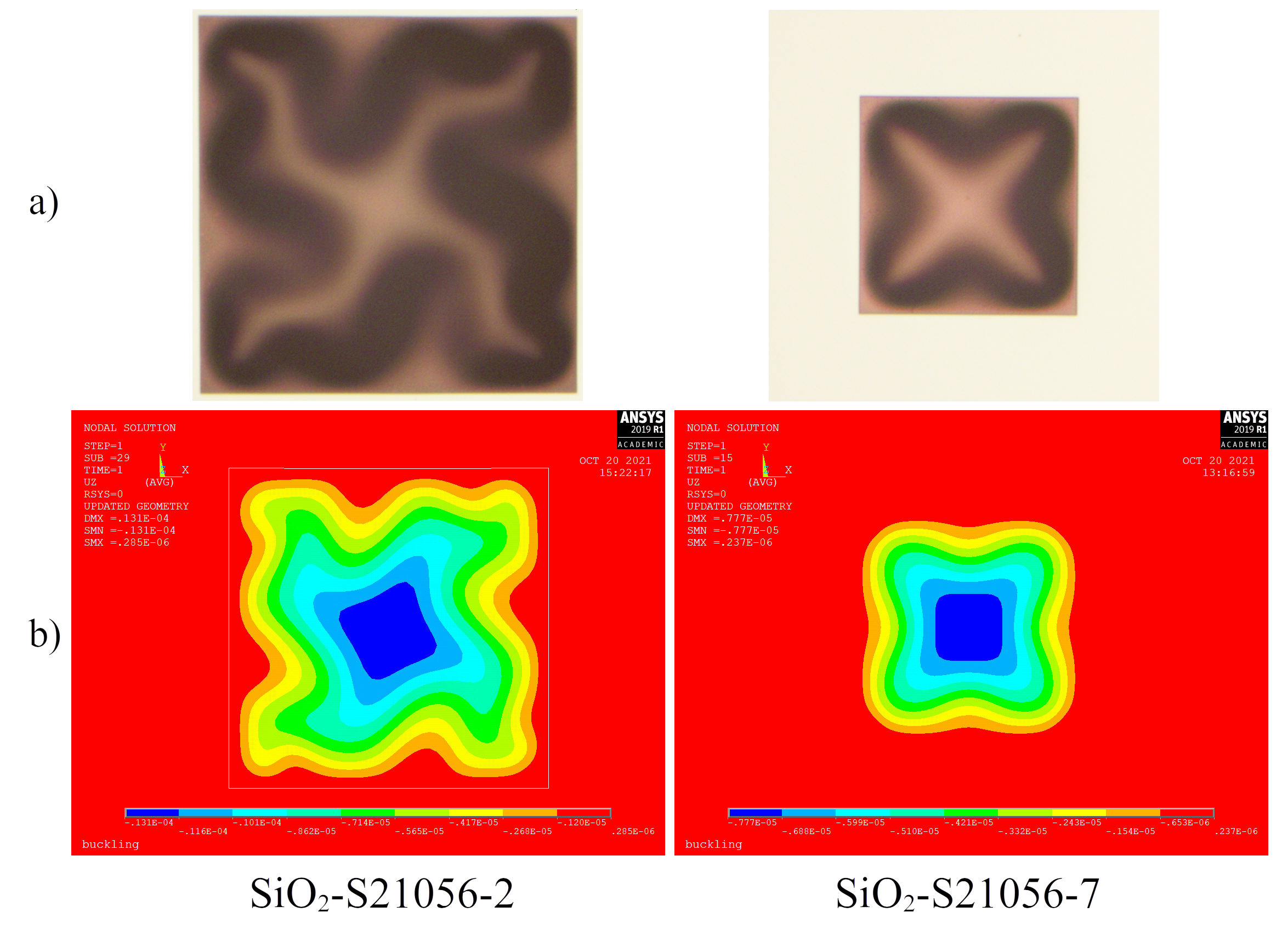}
\caption{\label{fig:photos_fem_600dpi} Comparison of top views of IBS $\mathrm{SiO_{2}}$ membranes S21056-2 (left) and S21056-7 (right): a) photos taken with an optical microscope, b) finite-element simulations.}
\end{center}
\end{figure}

\begin{figure*}
\begin{center}
\includegraphics[scale = 0.25]{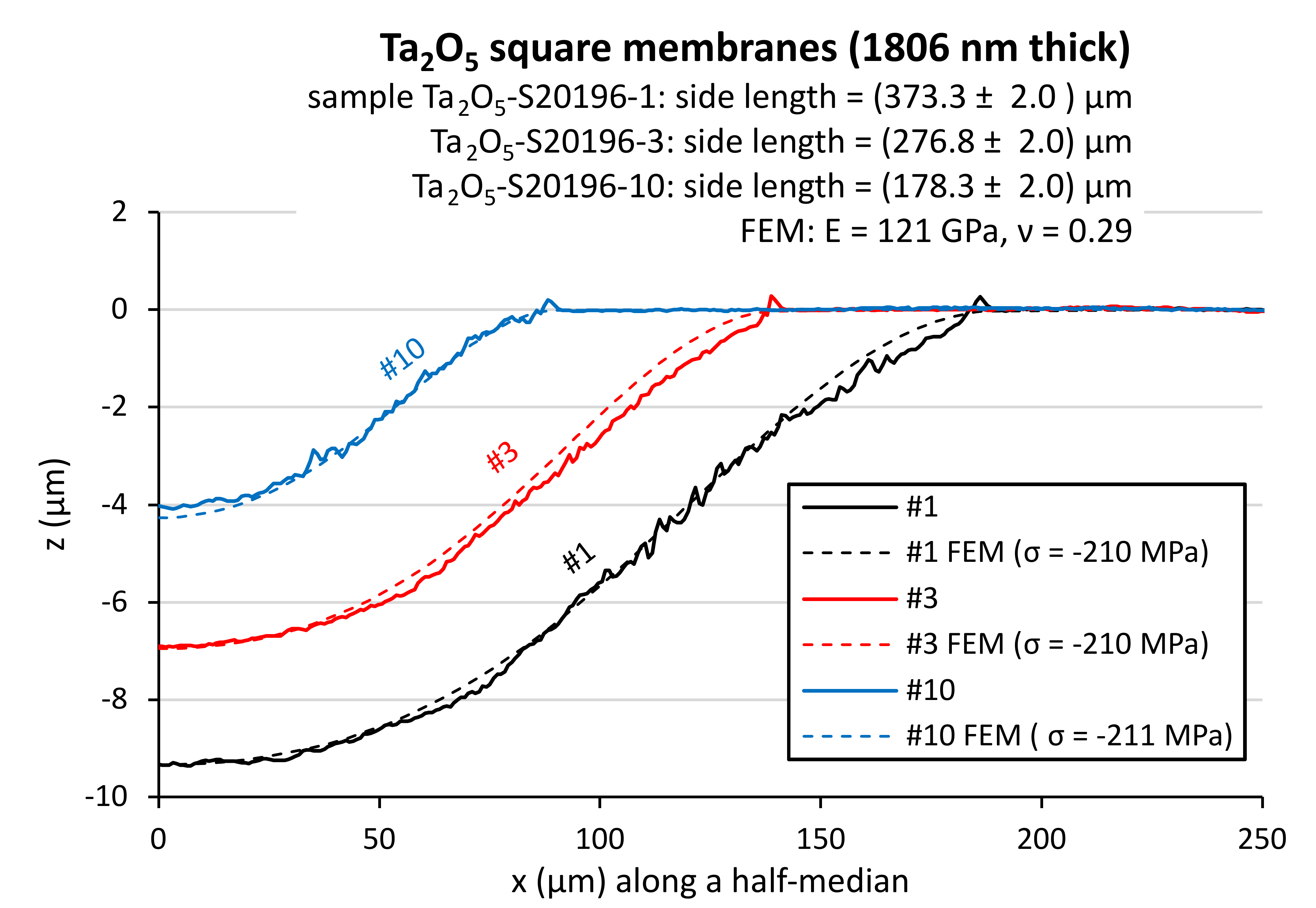}
\caption{\label{fig:fig_ta2o5_s20196_2d_profiles}Measured and simulated deflections for 1806 nm thick IBS $\mathrm{Ta_{2}O_{5}}$ membranes.}
\end{center}
\end{figure*}

\begin{figure*}
\begin{center}
\includegraphics[scale = 0.25]{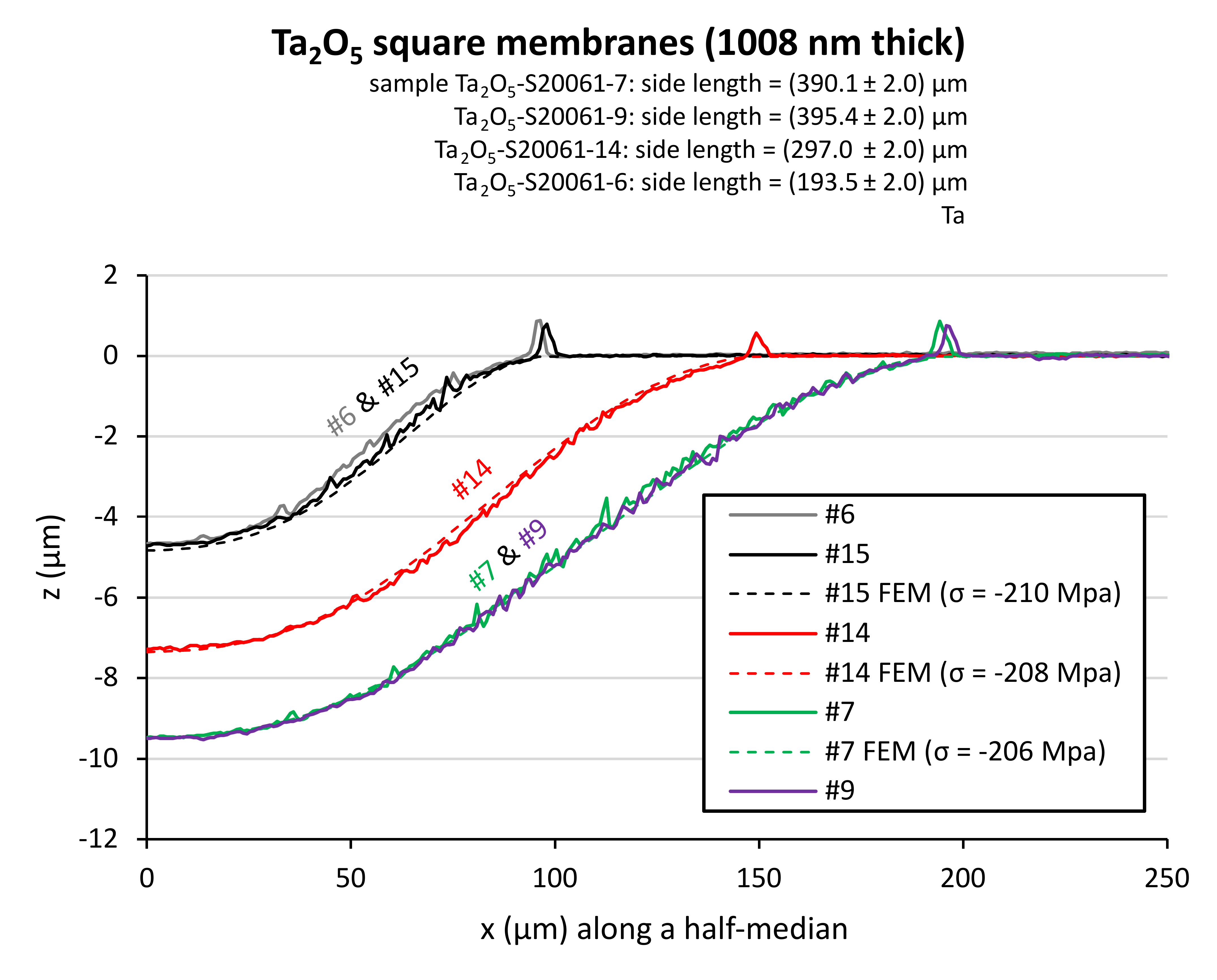}
\caption{\label{fig:fig_ta2o5_s20061_2d_profiles}Measured and simulated deflections for 1008 nm thick IBS $\mathrm{Ta_{2}O_{5}}$ membranes.}
\end{center}
\end{figure*}

\begin{figure*}
\begin{center}
\includegraphics[scale = 0.25]{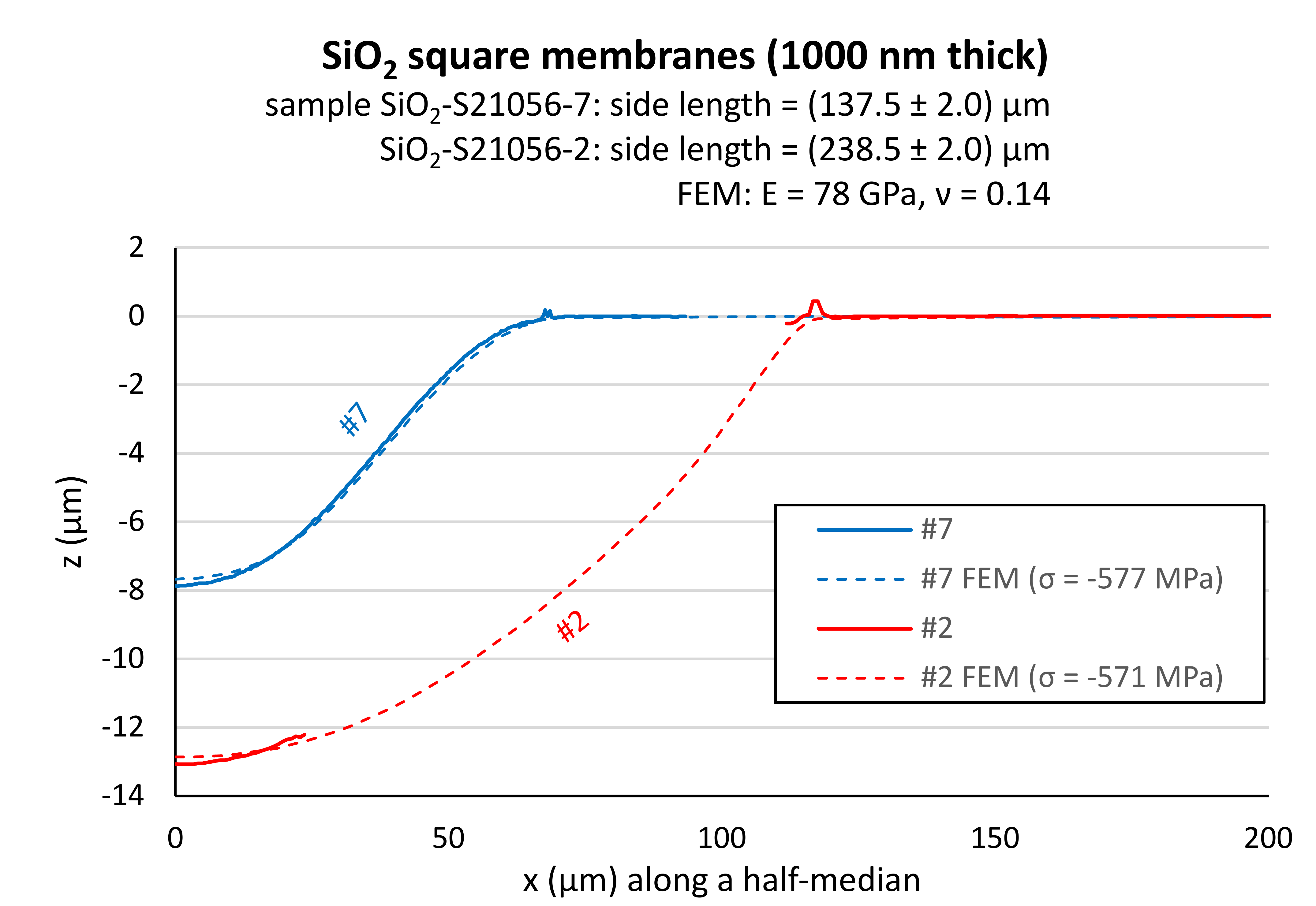}
\caption{\label{fig:fig_sio2_s21056_2d_profiles}Measured and simulated deflections for 1000 nm thick IBS $\mathrm{SiO_{2}}$ membranes.}
\end{center}
\end{figure*}

\begin{figure*}
\begin{center}
\includegraphics[scale = 0.25]{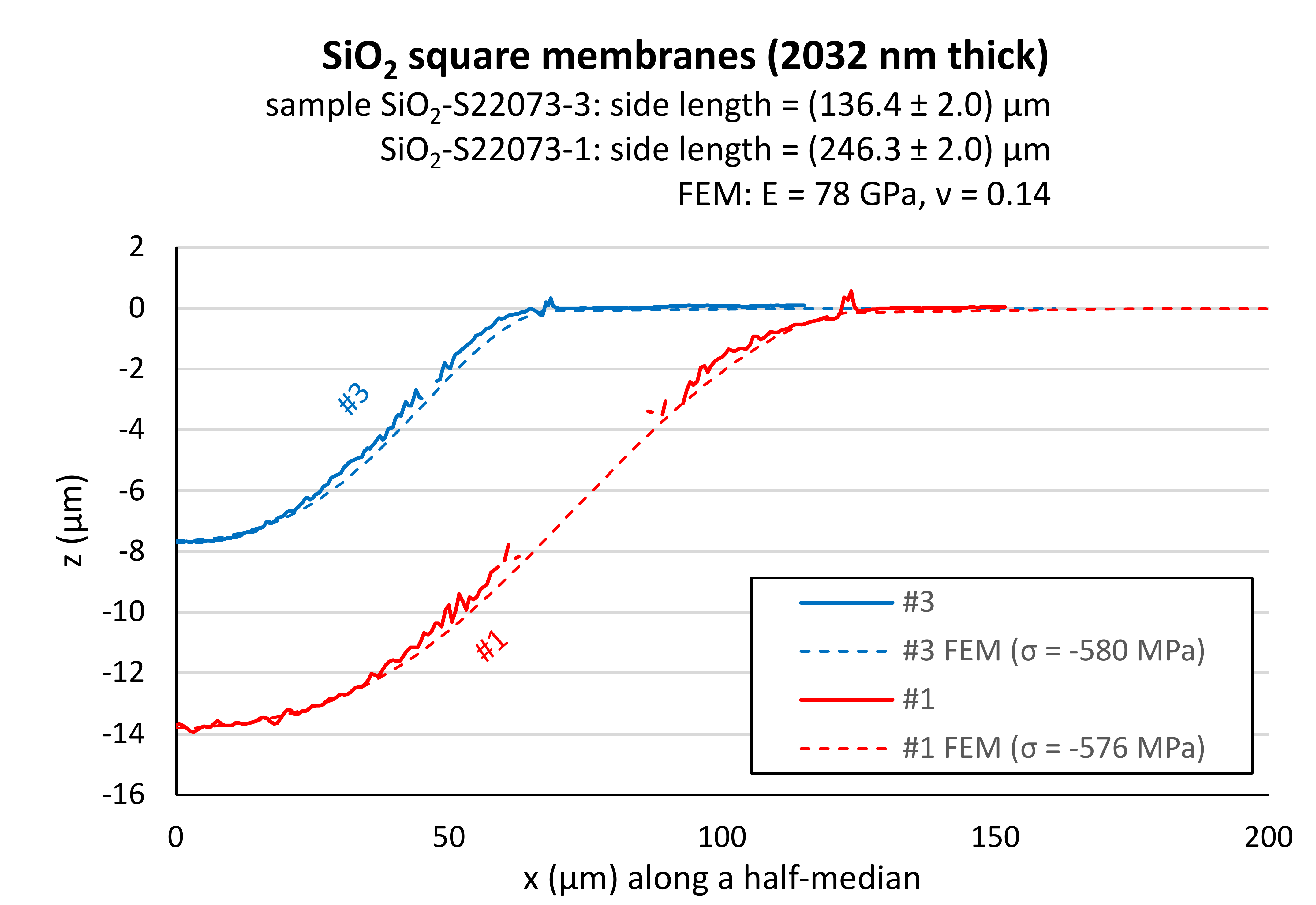}
\caption{\label{fig:fig_sio2_s22073_2d_profiles}Measured and simulated deflections for 2032 nm thick IBS $\mathrm{SiO_{2}}$ membranes.}
\end{center}
\end{figure*}

\begin{figure}[htbp]
\begin{center}
\includegraphics[scale = 0.25]{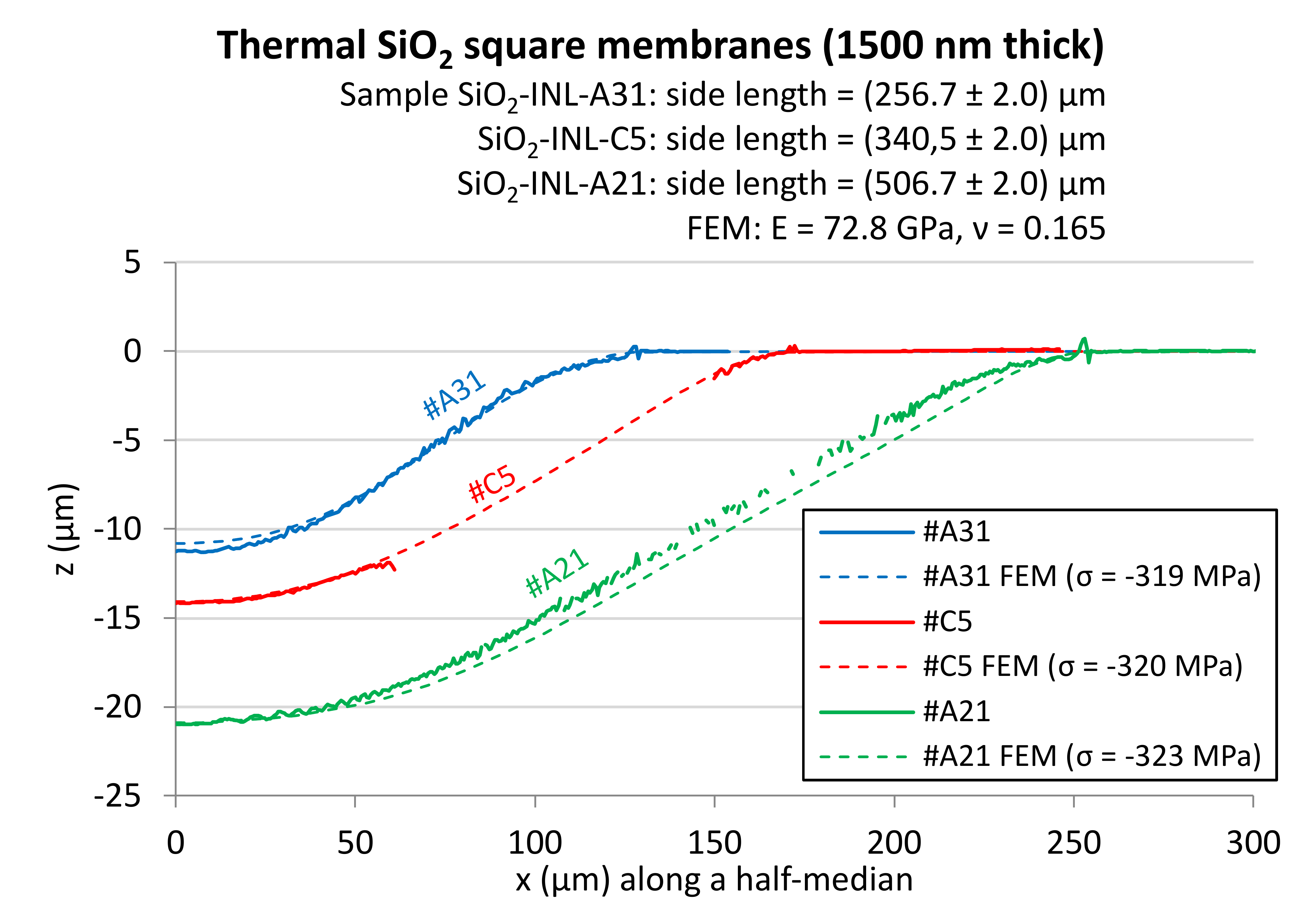}
\caption{\label{fig:fig_sio2_inl_2d_profiles}Measured and simulated deflections for 1500 nm thick thermal $\mathrm{SiO_{2}}$ membranes.}
\end{center}
\end{figure}

As an example, photographs of two IBS $\mathrm{SiO_{2}}$ membranes and the corresponding simulated profiles are shown in Fig. \ref{fig:photos_fem_600dpi}-a and \ref{fig:photos_fem_600dpi}-b, respectively. Observed and simulated deflection profiles, as probed along a half median, are compared in Figs. \ref{fig:fig_ta2o5_s20196_2d_profiles} and \ref{fig:fig_ta2o5_s20061_2d_profiles} for IBS $\mathrm{Ta_{2}O_{5}}$ membranes, and in Fig. \ref{fig:fig_sio2_s21056_2d_profiles} and \ref{fig:fig_sio2_s22073_2d_profiles} for IBS $\mathrm{SiO_{2}}$ membranes. Results for membranes of thermal $\mathrm{SiO_{2}}$ are presented in Fig. \ref{fig:fig_sio2_inl_2d_profiles}. Note that for some strongly deflected samples, some experimental data points are missing due to the maximum detectable slope limit set by the objective of the profilometer. 

The best-fit stress values used in the simulations are reported in the right column of Table \ref{tab:membranes_characteristics}. The stress state of the IBS $\mathrm{Ta_{2}O_{5}}$ membranes is about $-209$ MPa, whereas the IBS $\mathrm{SiO_{2}}$ membranes are in a higher compressive stress state of about $-576$ MPa. Such values seem to be independent from the sample thickness, as can be deduced from the comparison of IBS $\mathrm{Ta_{2}O_{5}}$ samples S20061 and S20196 and IBS $\mathrm{SiO_{2}}$ samples S21056 and S22073 with similar dimensions; they are also rather insensitive to the uncertainties about the Young's modulus and Poisson's ratio values used in the simulations. For the \SI{2032}{nm} thick IBS SiO$_2$ membranes, for example, additional simulations were performed with a 7\% lower Young's modulus and a 15\% higher Poisson's ratio (that is, as in bulk fused silica), yielding a change of the estimated stress level of only about 1\%: -572 MPa and -573 MPa for samples $\mathrm{SiO_{2}}$-S22073-1 and -3, respectively. This is analogous to the case of a drumhead, whose vibration frequency depends mainly on its tension and not on its rigidity.

The average stress in thermal $\mathrm{SiO_{2}}$ grown at 1130 $^{\circ}$C was found equal to $-321$ MPa, in good agreement with the literature \cite{Jaccodine_1966,Malhaire_1999}. In this case, the residual stress is mainly due to the mismatch of the thermal expansion coefficients of the thermally-grown $\mathrm{SiO_{2}}$ thin film and of the silicon substrate.

Assuming that the thin films' thickness and elastic constants are well known, the relative error on our stress evaluation method can be estimated to be less than 5\%, due to the uncertainties on the measurements of size and deflection of the membranes.

\section{Conclusions}
We developed an alternative method for the determination of compressive stresses in thin films, based on measurements and finite-element simulations of buckled self-supporting membranes, which appears to be a suitable alternative to the curvature method. As a large number of samples can be produced out of a single coated substrate, a high reliability and repeatability of the results can be obtained with our method; this also opens up the possibility of probing the stress uniformity across a large-surface sample, or of assessing whether stress is affected by the presence of a composition gradient. In addition, these very same membranes could be used for many other characterizations: micro-Raman spectroscopy, X-ray diffraction, ion-beam analyses (Rutherford back-scattering and elastic-recoil detection), etc.

Finally, we found that the average stress of IBS $\mathrm{Ta_{2}O_{5}}$ thin films is $-209$ MPa, and that IBS $\mathrm{SiO_{2}}$ thin films feature a higher average stress of $-576$ MPa. For comparison, the average stress in thermal $\mathrm{SiO_{2}}$ grown at 1130 $^{\circ}$C was found equal to $-321$ MPa, in good agreement with the literature.

\section{Acknowledgments}
The authors gratefully acknowledge the support of the French Agence Nationale de la Recherche (ANR) through Grant No. ANR-18-CE08-0023 (project ViSIONs) and of the national program Investments for the Future through Grant No. ANR-11-LABX-022-01 (managed by ANR). The membranes were processed at the Plateforme technologique Nanolyon of the Institut des Nanotechnologies de Lyon (INL). The authors thank J. Degouttes for his technical support on this project, and C. Gaillard for access to the chemistry laboratory at the Institut de Physique des deux Infinis de Lyon (IP2I).

\end{document}